\documentclass{ws-procs9x6-cpt19}

\usepackage{tikz}
\usetikzlibrary{positioning,fit,shapes.geometric,backgrounds,arrows,shapes}

\def\thpr{these proceedings}

\def\sb{\overline{s}{}}

\def\frac#1#2{{\textstyle{{#1}\over {#2}}}}

\def\ibar{{\mathrel{\rlap{$I$} \raise1pt\hbox{--}}}}

\def\lsim{\mathrel{\rlap{\lower4pt\hbox{\hskip1pt$\sim$}}
    \raise1pt\hbox{$<$}}}
\def\gsim{\mathrel{\rlap{\lower4pt\hbox{\hskip1pt$\sim$}}
    \raise1pt\hbox{$>$}}}
\def\sqr#1#2{{\vcenter{\vbox{\hrule height.#2pt
         \hbox{\vrule width.#2pt height#1pt \kern#1pt
         \vrule width.#2pt}
         \hrule height.#2pt}}}}

\def\etal{{\it et al.}}

\newcommand{\beq}{\begin{equation*}}
\newcommand{\eeq}{\end{equation*}}
\newcommand{\bea}{\begin{eqnarray}}
\newcommand{\eea}{\end{eqnarray}}
\newcommand{\bit}{\begin{itemize}}
\newcommand{\eit}{\end{itemize}}
\newcommand{\bcn}{\begin{center}}
\newcommand{\ecn}{\end{center}}

\begin{document}

\newcommand{\refeq}[1]{(\ref{#1})}
\def\etal {{\it et al.}}
%any other macros go here 

\title{Maximal Tests in Minimal Gravity}

\author{Jay D.\ Tasson}

\address{
  Physics and Astronomy Department, Carleton College,\\
  Northfield, MN 55057, USA\\
{\rm LIGO-P1900177}}

\begin{abstract}
Recent tests have generated impressive reach 
in the gravity sector of the Standard-Model Extension.
This contribution to the CPT'19 proceedings
summarizes this progress 
and maps the structure of work in the gravity sector.
\end{abstract}

\bodymatter

\section{Lorentz violation in gravity}
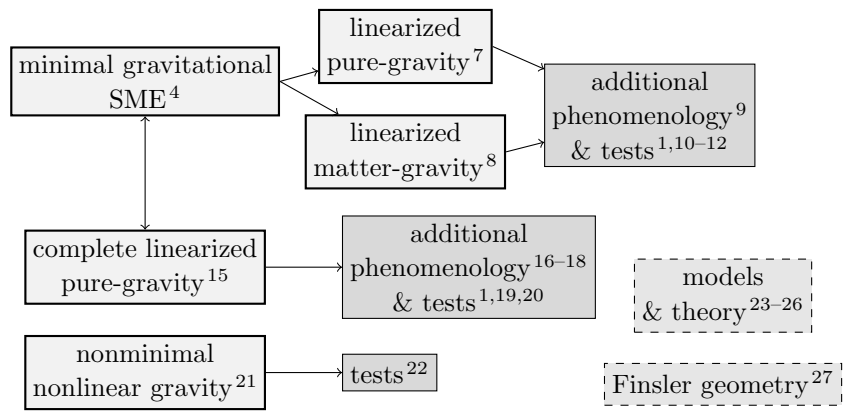
\begin{figure}
\begin{center}
\begin{tikzpicture}[%
    node distance = .4cm,
  inner sep=1mm,
  expt/.style={rectangle, trapezium left angle=120, trapezium right angle=60, thin, fill=gray!30, draw=black, align=center},
  action/.style={rectangle, trapezium left angle=120, trapezium right angle=60, thick, fill=gray!10, draw=black, align=center},
  oact/.style={rectangle, trapezium left angle=120, trapezium right angle=60, dashed, fill=gray!20, draw=black, align=center},
    loop/.style={ellipse, thick, fill=yellow!30, draw=yellow!50, align=center},
  title/.style={font=\LARGE\scshape,node distance=16pt, text=black!40, inner sep=1mm},
  background/.style={rectangle, rounded corners, fill=black!5, draw=black!15, inner sep=4mm}
]

\node[action] (akgrav)
       {minimal gravitational \\
         SME \cite{SME2}};
  \node[action] (lvpn) [above right = -0.5cm and 0.5cm of akgrav] {linearized\\
 pure-gravity \cite{lvpn}};
  \node[action] (lvgap) [ below =of lvpn] {linearized\\
 matter-gravity \cite{lvgap}};
  \node[expt] (exp) [above right = -0.7cm and 0.5cm of lvgap] {
        additional\\ phenomenology\cite{phenom}\\
\&
        tests\cite{datatables, mo,lpl,trostel}};
  \node[action] (akmm) [ below =1.5cm of akgrav] {complete linearized\\
         pure-gravity \cite{mkgrav}};
  \node[expt] (nonmin) [ right= 1cm of akmm] {additional\\ phenomenology \cite{mmkgrav,phenom2,phenom3}\\
\&
        tests \cite{datatables,srexpt,psrnm}};
  \node[action] (nonlin) [ below  = of akmm] {nonminimal \\ nonlinear gravity \cite{qb16}};

  \node[expt] (nlx) [ right= 1cm of nonlin] {tests \cite{psrnl}};

  \node[oact] (other) [below right = -0.8cm and 0.5cm of nonmin] {models \\ \& theory \cite{dc,cl,rb,thry}};

  \node[oact] (fin) [below = of other] {Finsler geometry\cite{be}};
  
  \node[fit=(akgrav)] (chart) {};
  
  \draw[->] (akgrav.east) -- (lvpn);
  \draw[->] (akgrav.east) -- (lvgap);
  \draw[->] (lvpn.east) -- (exp);
  \draw[->] (lvgap.east) -- (exp);
  \draw[<->] (akgrav.south) -- (akmm);
  \draw[->] (akmm.east) -- (nonmin);
  \draw[->] (nonlin.east) -- (nlx);
  
\end{tikzpicture}
\end{center}
\caption{Progress in SME gravity as of CPT'19.
Light gray boxes show the various limits of the gravity sector 
that have been explored.
Dark gray boxes show work 
that builds out the search in the respective limits.
Dashed boxes show theoretical contributions.} 
\label{aba:fig1}
\end{figure}

As demonstrated by the breadth of contributions to these proceedings 
and the ongoing growth of the 
{\it Data Tables for Lorentz and CPT Violation},\cite{datatables}
the search for Lorentz violation as a signal of new physics,
such as that originating at the Planck Scale,\cite{ks}
is an active research area.
The gravitational Standard-Model Extension (SME)\cite{SME1,SME2,SME3}
provides a field-theory-based framework 
for performing the search systematically.
The structure of the SME can be thought of 
as a series expansion about known physics,
with additional terms of increasing mass dimension 
constructed from conventional fields
coupled to coefficients for Lorentz violation.\cite{jtmd}
The leading terms,
associated with operators of mass dimension $d=3,4$,
are known as the minimal SME.
In the gravity sector,
phenomenology has been developed
and tests have been performed
based on a variety of complementary limits
of the full SME.
Relations among these efforts
are summarized graphically in Fig.\ \ref{aba:fig1}.

The framework for phenomenology in the gravity sector of the SME 
began in 2004 with Ref.\ \refcite{SME2},
which developed the Lagrange density
and associated theory to be used 
in searches for minimal Lorentz violation in gravity.
Lorentz-violating effects in gravity
can be understood as coming from the pure-gravity sector
through Lorentz-violating modifications 
to the dynamics of the gravitational field, \cite{lvpn}
or through gravitational couplings in Lorentz-violating terms 
in the other sectors of the theory.\cite{lvgap}
In the latter case,
Lorentz-violating effects are dependent on the species of matter 
contained in the test and source bodies,
while in the former case they are not.
References \refcite{lvpn,lvgap} address theory and phenomenology 
associated with minimal terms in pure gravity and matter-gravity couplings, respectively.
A large amount of additional phenomenology\cite{phenom} 
and experimental and observational searches\cite{datatables}
have been done based on these works,  
some of which are discussed in Sec.\ \ref{reach} 
and elsewhere in these proceedings.\cite{mo,lpl,trostel}

Some nonminimal gravity-sector terms
were analyzed for short-range gravity experiments\cite{bkx} 
and for gravitational \v Cerenkov radiation,\cite{cer}
and the complete linearized theory of pure gravity 
was developed in Ref.\ \refcite{mkgrav},
with initial applications to gravitational waves (GWs).
Since then,
additional phenomenology\cite{mmkgrav,phenom2,phenom3} 
as well as experimental and observational work\cite{datatables} 
has been done in nonminimal gravity.
Examples of searches 
in nonminimal gravity 
are contained in these proceedings.\cite{trostel,srexpt,psrnm} 
We note in passing the expected overlap 
between the linearized limit of the minimal work of Refs.\ \refcite{SME2,lvpn}
and the minimal limit of the complete linearized theory in Ref.\ \refcite{mkgrav}.
Study of the nonminimal gravity sector 
beyond the linearized limit has also begun.\cite{qb16,psrnl}

In addition to work aimed directly 
at seeking signals of Lorentz violation in experiments,
several theory-oriented results deserve discussion in this context.
While it is difficult to capture the volume of work done in this area
in this short summary,
examples discussed in these proceedings
include exploration of specific Lorentz-violating models 
that generate nonzero SME coefficients\cite{dc,cl}
and the implications of geometric constraints on Lorentz violation.\cite{rb}
The question of geometric constraints 
has also inspired consideration of Finsler geometry
as a geometric framework for Lorentz violation.\cite{be}

\section{Maximal reach}
\label{reach}

Several recent and ongoing efforts 
have improved sensitivities to Lorentz violation 
in the minimal gravitational sector,
or are expected to do so in the near future.
A number of these are discussed elsewhere in these proceedings
including improved sensitivities through matter--gravity couplings
based on an analysis of data from the MICROSCOPE mission,\cite{mo}
results from the analysis of solar-system data,\cite{lpl}
and tests based on interferometric gyroscopes.\cite{trostel,adv}
Significant improvements in the laboratory
were also achieved using gravimeters.\cite{gravimeter}
In this section,
we summarize the recent effort providing the greatest reach,
multimessenger astronomy.

On August 17, 2017,
GWs and photons from the same astrophysical event
were observed for the first time.\cite{gwgrb}
A gamma-ray burst arrived $(1.74 \pm 0.05)\,$s after
the GWs from the coalescence of a pair of neutron stars.
This observation, 
along with modeling suggesting up to a few seconds of lag 
between the coalescence and gamma-rays emission,
led to a best-ever comparison of the speed of GWs and light.
Such tests provide a particularly sensitive probe
of $d=4$ SME gravity coefficients
due to the long propagation distance involved
and because GW tests
based on birefringence\cite{mkgrav,phenom2} 
and/or dispersion,\cite{mkgrav,phenom2,as}
while powerful for $d > 4$,
are insensitive to $d=4$ coefficients.
Using a maximum-reach analysis,
in which the nine minimal $\sb^{(4)}_{jk}$ gravity-sector coefficients 
are taken as nonzero one at time,
the reach for  all nine coefficients was improved over prior limits,
most of which came from the analysis of \v Cerenkov radiation 
by cosmic rays.\cite{cer}
The upper bound on the isotropic $\sb^{(4)}_{00}$ coefficient
was inaccessible to \v Cerenkov constraints,
hence an improvement of ten orders of magnitude
was achieved here,
while improvements of up to a factor of 40 
were achieved for the other coefficients.
Future observations of multimessenger events
offer several avenues of improvement.
Events further away will improve the overall sensitivity,
at least nine events distributed across the sky will enable
the estimation of all nine $\sb^{(4)}_{jk}$ coefficients together,
and events at a variety of distances will
disentangle speed differences from emission-time differences.
The future is bright for seeking Lorentz violation
with GWs.

\section*{Acknowledgments}
J.T.\ is supported by NSF grant PHY1806990 to Carleton College.

\noindent
Note: the additional length of this document
relative to the published version
is due to the addition of arXiv numbers in the references
for papers in the proceedings volume.

\end{document}